# ENERGY EFFICIENT CLUSTERING AND ROUTING IN MOBILE WIRELESS SENSOR NETWORK

Getsy S Sara[1] Kalaiarasi.R[2], Neelavathy Pari.S[3] and Sridharan .D[4]

Department of Electronics & Communication Engineering, College of Engineering, Anna University
Chennai, India, 600 025.
[1]getsysudhir@gmail.com, [2] rajendranarasi@gmail.com [3]neela_pari@yahoo.com,
[4]sridhar@annauniv.edu.

***ABSTRACT***

*A critical need in Mobile Wireless Sensor Network (MWSN) is to achieve energy efficiency during routing as the sensor nodes have scarce energy resource. The nodes' mobility in MWSN poses a challenge to design an energy efficient routing protocol. Clustering helps to achieve energy efficiency by reducing the organization complexity overhead of the network which is proportional to the number of nodes in the network. This paper proposes a novel hybrid multipath routing algorithm with an efficient clustering technique. A node is selected as cluster head if it has high surplus energy, better transmission range and least mobility. The Energy Aware (EA) selection mechanism and the Maximal Nodal Surplus Energy estimation technique incorporated in this algorithm improves the energy performance during routing. Simulation results can show that the proposed clustering and routing algorithm can scale well in dynamic and energy deficient mobile sensor network.*

## 1. INTRODUCTION

Recent researches have shown that introducing mobility in wireless sensor network is advantageous as the mobile nodes can relocate after initial deployment to achieve the desired density requirement and reduce the energy holes in the network thereby increasing the network life time. A mobile wireless sensor network consists of tiny sensor nodes which has three basic components: a sensing subsystem for data acquisition from the physical surrounding environment, a processing subsystem for local data processing and storage, a wireless communication subsystem for data transmission [16]. Mobility of these nodes can be achieved by equipping the sensor nodes with mobilizers, springs[17] and wheels [18] or they can be attached to transporters like vehicles, animals, robots [19] etc. Mobility in wireless sensor network is also very taxing [2] as path breakage happens frequently due to node movement. Frequent location updates from a mobile node is needed to establish routing which eventually leads to excessive drain of sensor node's battery supply and also increases collisions [20]. Lifetime of a sensor network depends on energy supply. Therefore it is necessary to design energy efficient routing protocol.

This paper is distributed as follows – Section 2 gives an idea on prior works done in this area. Section 3 gives details about Energy Efficient Clustering and Routing Protocol in Mobile Wireless Sensor Network. The simulation of the proposed routing protocol is given in Section 4. Finally, Section 5 summarizes this paper.

## 2. RELATED WORK

Several researchers have focused to provide very energy efficient routing protocols for Wireless Sensor Network with mobile sinks . Kisuk kweon et al have proposed the Grid Based Energy Efficient Routing (GBEER) [5] for communication from multiple sources to multiple mobile sinks in wireless sensor network. With the global location information a permanent grid structure is built. Data requests are routed to the source along the grid and data is sent back to the sinks. The grid quorum solution is adopted to effectively advertise and request the data for mobile sinks. The communication overhead caused by sink's mobility is limited to the grid cell. There is no additional energy consumption due to multiple events because only one grid structure is built independently of the event. Zhi-Feng Duan etal designed a three layer mobile node architecture [6] to organize all sensors in MWSN. Here the Shortest Path (SP) routing protocol is used to adapt sensors to update the network topology. SP provides an elegant solution to node movement in multilayer MWSN and reduces energy dissipation. In [3] and [7], Xiaoxia Huang proposed a robust cooperative routing protocol based on cross layer design with MAC layer as the anchor, operated under IEEE 802.11 MAC protocol. Robustness against path breakage is improved here thereby improving the energy efficiency of the network. Chuan – Ming etal [21] proposed distributed clustering algorithm for data gathering in mobile wireless sensor network. The cluster formation was done using Cluster with Mobility mechanism (CM). Cluster head was elected using two distributed algorithms. It was observed that a better clustering factor and lesser energy consumption were achieved.

## 3. ENERGY EFFICIENT CLUSTERING AND ROUTING PROTOCOL

The proposed algorithm is an extension of our work in [4]. The idea of this algorithm is to design multipath hybrid routing protocol ($E^2RP$) with efficient clustering technique ($E^2C$) to achieve less dissipation of energy in mobile sensor network.

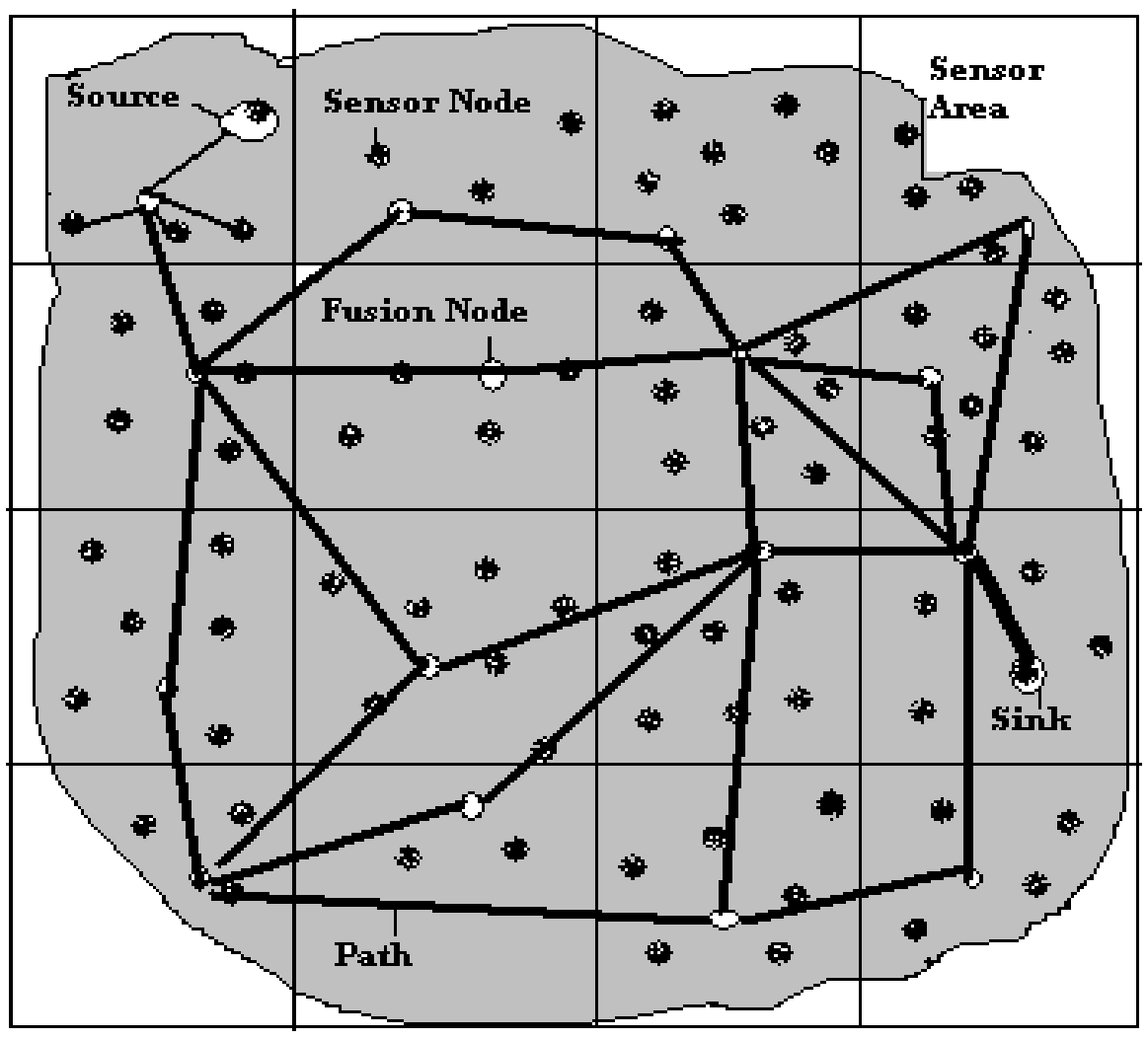

Figure 1. Formation of precincts in MWSN

### 3.1. Energy Efficient Clustering

The entire sensor network area is assumed to be circumscribed into a big square and then divided into different square zones. The square zone is chosen here as it allows union of zones without having holes and also simplifies the design of clustering algorithm [10].Assume that each node can detect the signal strength within its radio range. The gateway nodes are the nodes within a precinct that have at least one neighbouring precinct node within its communication range.

Let n be the average number of sensors in the network. Initially it is assumed that all the nodes have energy of 5J. The fusion nodes for each precinct are selected based on the combination of several metrics such as low node mobility, high surplus energy and best transmission range [22]. All the nodes in the precinct have the same chance to become fusion node.

### 3.1.1. Formation of Fusion Node

All the nodes in the precinct broadcast a HELLO packet which consists of its surplus energy and its probability to become the fusion head which is calculated based on node's mobility metric, transmission range and surplus energy. The virtual ID (VID) [12] for each sensor node is created as,

$$\text{VID} = \text{surplus energy}/\text{No. of nodes in the precinct} \qquad (1)$$

### 3.1.2. Estimation of Surplus Energy

A sensor node consumes energy when it is sensing, generating data, receiving, transmitting or in standby mode. Let $e_g$ be the power used for sensing and generating one bit of data and $e_s$ be the standby power of each node. The $e_g$ and $e_s$ value is assumed to be same for all the nodes. A node needs $E_{elec}$ = 50 nJ to run the circuitry and $E_{amp}$ = 100 pJ/bit/m2 for transmitting amplifier [11].

Power consumed for receiving one bit of data, $e_r = E_{elec}$

Power consumed for transmitting one bit of data to a neighboring node j is given by

$$e_{ij} = E_{elec} + E_{amp} \,.d_{ij}^{\alpha} \qquad (2)$$

where

$\alpha$ - path loss component

$d_{ij}^{\alpha}$ – Euclidean distance of node i and node j respectively

Assuming each node i has initial battery energy $E_i$, the uniformed mean power consumption of node i is calculated as

$$w_i = [e_s + e_g \,.\, r_i + e_{r.}\, r_j + r_{i.}\, e_{ij}\,] \,/\, E_i \qquad (3)$$

where

$r_i$ & $r_j$ – traffic generating rate at node i and node j respectively

$$\text{Surplus energy}, \; E_s = E_i - w_i \qquad (4)$$

### 3.1.3. Estimation of Transmission Range of a Sensor Node

Using Friss free space formula, the transmission range of a sensor node is calculated [23] as

$$R_{max} = [\lambda/4\pi]*[\sqrt{(P_t *G_t *G_r * (1-|\Gamma|^2) / P_r)}] \qquad (5)$$

where

$\lambda$ – Operating wavelength

$P_t$ - power transmitted by the sensor

$P_r$ – receiver sensitivity

$G_t$ – Gain of transmitting antenna

$G_r$ – Gain of receiving antenna

$|\Gamma|^2$ – reflected power coefficient of receiving antenna.

### 3.1.4. Estimation of mobility of a Sensor Node

The mobility metric is calculated as the measure of relative motion of nodes. Mobility measure is normalized by the number of nodes and the continuous functions of time that represent the quantitative measures of relative motion between nodes at time t [24].

$$M(t) = (1/N) \sum_{i=0}^{N-1} M_i(t) \qquad (6)$$

where

$M_i(t)$ – relative movement of other nodes as seen by node i
N – number of sensor nodes in a precinct

$$M_i(t) = (1/N-1) \sum_{j=0}^{N-1} |d'_{ij}(t)| \qquad (7)$$

where

$d_{ij}(t)$ – distance between node I and node j at time t
$d'_{ij}(t)$ – time derivative of $d_{ij}(t)$

### 3.1.5. Calculation of probability of a Sensor Node becoming Fusion node

Let $P(F_i)$ denote the probability of a node becoming the best Fusion Head. Let us consider there are N numbers of nodes in the precinct. Probability of a node becoming the precinct head is 1/N initially. $E_{th}$, $R_{th}$, $M_{th}$ are the threshold values of surplus energy, transmission range and mobility respectively. Let $E_i$ denote the event that the node i has residual energy greater than $E_{th}$. $R_i$ denote the event that the node i has transmission range greater than $R_{th}$. Let $M_i$ denote the event that the node i's mobility metric is lesser than $M_{th}$.

Using Baye's Rule [25],

$$P(F_i) = P(E_i)^* P(F_i/E_i) + P(R_i)^* P(F_i/R_i) + P(M_i)^* P(F_i/M_i) \qquad (8)$$

The node with the highest $P(F_i)$ is chosen as the fusion head. Periodically, the $P(F_i)$ value is checked. If the $P(F_i) < P_{th}(F_i)$, then the fusion node sets its VID as 0. A non fusion node which has the highest $P(F_i)$ at that time is chosen as next fusion head. $P_{th}(F_i)$ is chosen as 0.5.

## 3.2. Energy Efficient Routing

### 3.2.1. IntrA Precinct routing

The source node on detecting an event generates the Data Announcement packet (DA packet). DA packet consists of node's VID and data generation time. Many of the sensor nodes in a precinct may detect the event simultaneously and send the DA packet to the fusion node using IntrA Precinct routing. The Transmission Range threshold ($R_{th}$) ensures that every sensor node can communicate with the fusion node using single hop communication. The fusion node aggregates and compresses these packets using data fusion algorithm [5]. Then it checks if the destination is within its precinct. If so, proactively the event is send to the destination.

### 3.2.2. IntEr Precinct routing

The data is forwarded by the fusion node via the gateway nodes to the other precincts using the multipath reactive routing technique called Inter Precinct routing. Rather than choosing all the available multipaths, few of the best paths are selected based on the maximal nodal surplus energy. The RREQ message contains the following fields:

< Source address, source precinct id, sequence no., broadcast id, hop count, destination address, maximum surplus energy>

The RREP message contains the subsequent fields:

<Source address, destination address, destination precinct id, sequence number, hop count, readiness factor, maximum surplus energy, lifetime>.

If the readiness factor denotes 'Discard', a route error (RERR) message is propagated in the reverse path instead of RREP [4].

### 3.2.3. Energy aware selection mechanism

Each fusion node monitors its energy consumption and computes the energy drain rate ($DR_i$) [26] for every T second sampling interval by averaging the amount of energy consumption and estimating the energy dissipation per second during the past T seconds.

Drain rate of node i,

$$DR_i = \alpha * DR_{old} + (1-\alpha) * DR_{sample} \qquad (9)$$

where

$\alpha$ – constant

$DR_{old}$ – previous drain rate

$DR_{sample}$ - Current sample drain rate

Life time of a node i

$$L_i = E_s / DR_i \qquad (10)$$

The ratio of surplus energy ($E_s$) at node i to the drain rate indicates how long node i can keep up with the routing operations under current traffic conditions based on surplus energy. The readiness selection of a fusion node is based on battery capacity and predicted lifetime of a node [15]. The heuristic used to associate readiness are ('Discard', 'Moderate', 'High') to a pair (battery capacity, lifetime). This mechanism permits better load balancing to be obtained. It is assumed that if the surplus energy is less than 1J then it denotes low battery value. If the lifetime of a node is less than 10 seconds then it has a short lifetime and if life time is greater than 100 seconds then the node has a long life time.

### 3.2.4. Finding the Maximal Nodal Surplus Energy

When the intermediate fusion node receives a RREQ message from a gateway node, it checks if the sequence number specified in the RREQ message is greater than the node's sequence number [13]. If so, it compares the surplus energy in the RREQ message and the surplus energy of the node. In case the node's surplus energy is greater than that specified in the RREQ message, the surplus energy variable in the RREQ message is updated with the nodal surplus energy. By this method it is possible to achieve the value of maximum surplus energy among all fusion nodes in the specified path. The reverse paths are set up as the RREQ travels from a source to various destinations.

### 3.2.5. Sorting Multipath and Forwarding Data Packets

The paths in the route list are sorted by the descending value of surplus energy. The path with the maximum surplus energy is chosen to forward the data packets. Once the source node receives the RREP message containing the new path with maximum surplus energy, it forwards the data packets through this path.

The structure of the routing table at each node is shown in the figure 2.The advertised hop count of a node i for a destination represent the maximum hop count of the multiple paths for destination available at i. If the maximum hop count is considered, then the advertised hop count can never change for the same sequence number. The protocol only allows accepting alternate routes with lower hop count. This invariance is necessary to guarantee loop freedom [9].

| |
|---|
| Destination |
| Sequence Number |
| Advertised-hop count |
| Route list<br>{ next hop1, hopcount1, readiness( ) , max-surplus energy1}<br>{ next hop3, hopcount2, readiness( ) , max-surplus energy2}<br>.<br>.<br>{ next hop n, hop count n, readiness n ( ) , max-surplus energy n} |
| Expiration time |

Figure. 2. Structure of routing table entries at the fusion node for $E^2$MWSNRP

## 4. SIMULATION METHODOLOGY AND PERFORMANCE STUDY

The OMNET++ simulator is used [14] for implementing this algorithm. A dense sensor network of 100 nodes is being simulated in a field with 25*25 $m^2$ area. Each simulation last for 500 seconds. The random waypoint mobility model with a pause time of 30 seconds is used to simulate node movement. The mobile sensor node speed is set to be between 5m/sec and 20m/sec. For radio power consumption setting, $E_{elec}$ = 50 nJ/bit, $E_{amp}$ = 100 pJ/bit/m2 and a path loss exponent $\alpha$ =2 is set [11]. Each sensor node is assumed to have an initial energy of 5J. The transmitting power and the receiving power of each sensor node are presumed to be 0.66W and 0.395W respectively [11]. The $e_g$ and $e_s$ value is taken as 50 mW and 28.36 mW respectively and the radio range of a sensor node is 250m. Gt and Gr values are chosen as 1.2 dBi. A is assumed to be 0.3 and $\lambda$ is 900 MHz. The data is generated at a rate of 1Kbps. The distributed coordination function (DCF) of IEEE 802.11 is used as MAC layer. Each control packet is 36 bytes long and data packet is 64 bytes long.

To evaluate the performance of this algorithm, two key performance metrics are assessed; packet delivery fraction and network lifetime. Packet delivery fraction is defined as the ratio of data packets delivered to the destination to those generated by the source. The network lifetime is defined as the duration from the beginning of the simulation to the first time a node runs out of energy [4].

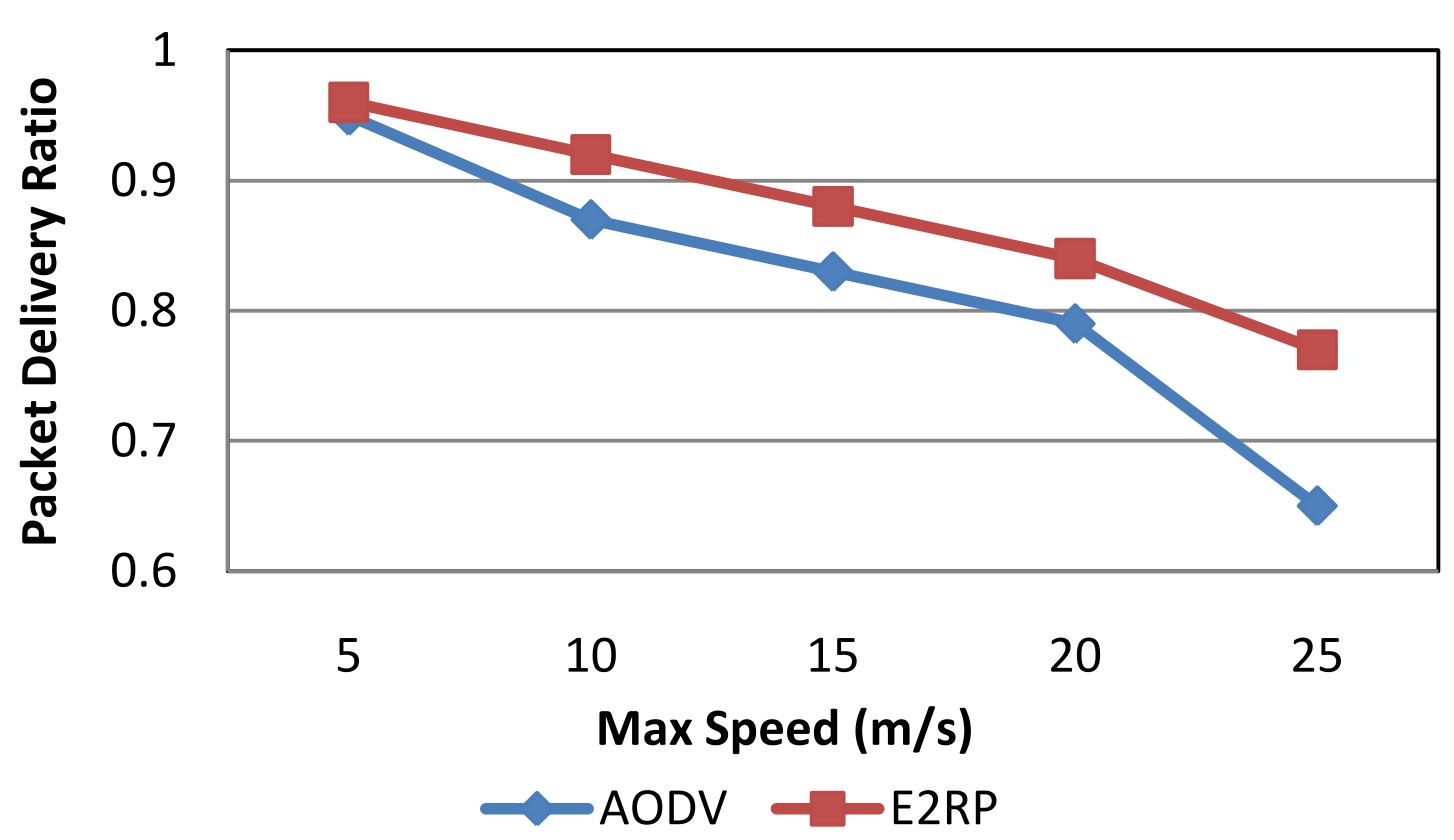

Figure.3. Increase in packet delivery ratio using $E^2$RP as compared to AODV

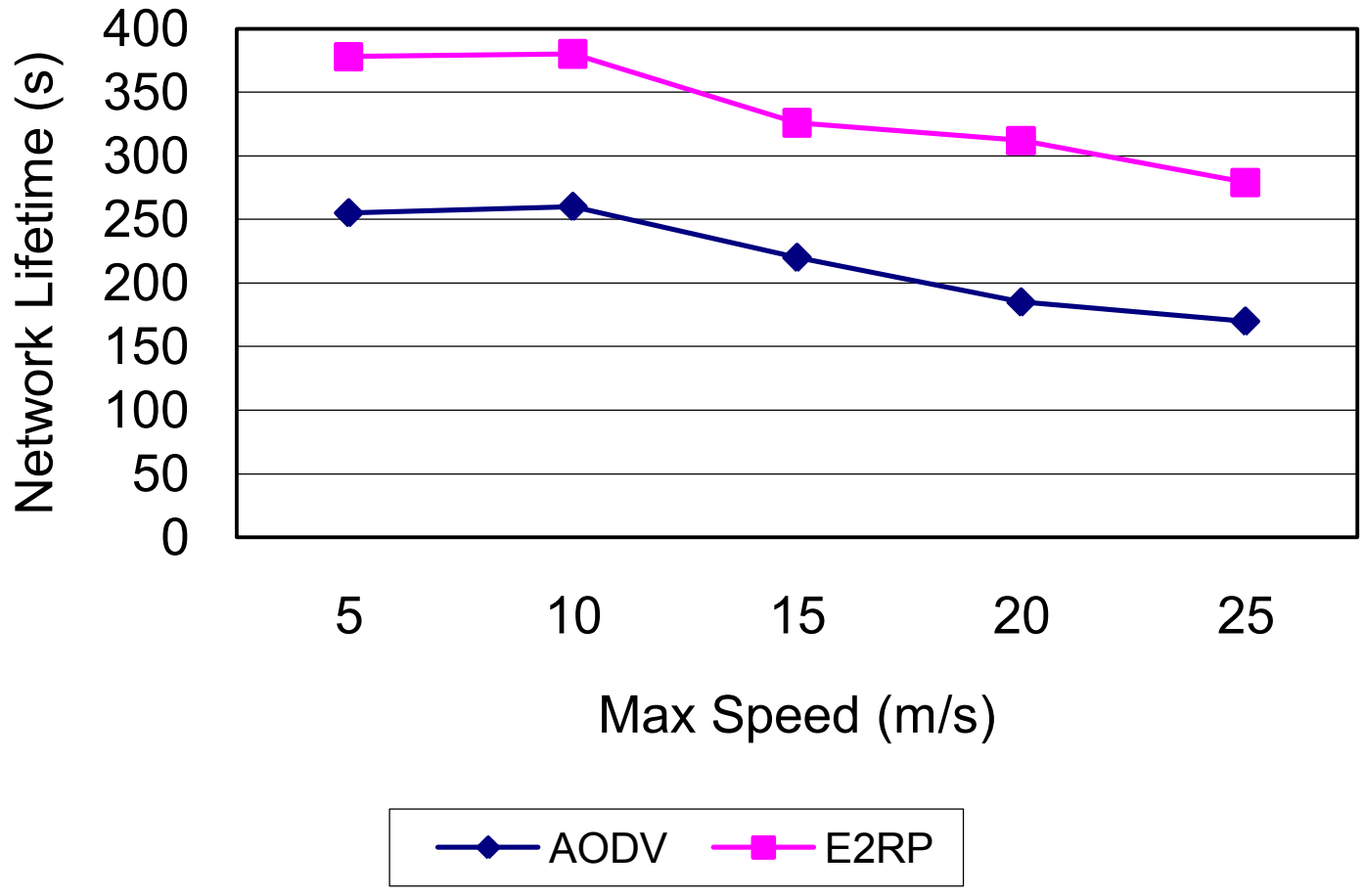

Figure.4. Increase in network life time using $E^2RP$ as compared to AODV

The performance of the routing protocol was simulated and compared with AODV routing protocol. It is observedfrom figure 3 and figure 4 that due to EA selection mechanism and the selection of route with maximal nodal surplus energy, the $E^2RP$ performed better than AODV in terms of packet delivery ratio and network life time. This algorithm will not stress sensor nodes with less residual energy thereby preventing the critical nodes from depleting their energy earlier and avoids route rediscovery for every route break [4].The proposed clustering technique is yet to be simulated and the improvement of the proposed routing algorithm along with the clustering algorithm is to be analysed. We expect a better performance in terms of lesser energy consumption because the lifespan of the fusion node is increased.

## 5. Conclusion

This paper presents an innovative and energy efficient clustering and routing protocol. The basic idea behind this algorithm is that the fusion head is selected based on Baye's rule which chooses the node with the highest surplus energy, least mobility and best transmission range. The routing protocol incorporates the hybrid routing concept. It uses the energy aware selection mechanism to choose the fusion nodes to route the data to the destination node. The route list consists of multipath with maximal nodal residual energy. In case of link failure and route breaks, new route discovery can be avoided. The proposed algorithm guarantees a longer network lifetime and better packet delivery ratio with less energy consumption.

## Acknowledgements

This paper is supported by the Junior Research Fellowship for Engineering and Technology under University Grants Commission India.

**Authors**

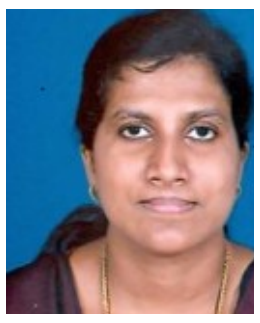

**Getsy S Sara** is currently pursuing her Ph.D degree at Anna University Chennai, India. She received her B.E degree with distinction in Electronics & Communication from Bharathiar University, India in 2004 and M.E degree with distinction in Digital Communication Engineering from Anna University, India in 2006. Her research interests include wireless ad hoc networking, sensor networks, energy efficient routing protocols and communication systems.

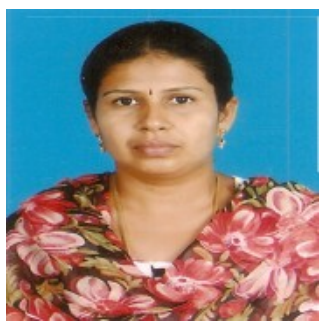

**Kalaiarasi R** is currently doing her Ph.D degree at Anna University Chennai, India. She received her B.Sc degree in Computer Science from Bharathiar University, India in 1996 and Master of Computer Applications from University of Madras in 2004. Her research interests include MANET, sensor network, mobile computing and network security.

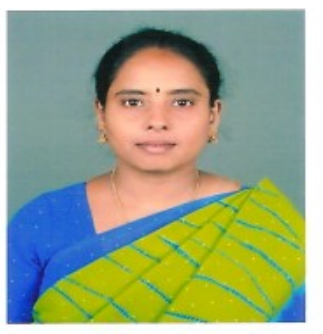

**Neelavathy Pari** received her B.E degree in Computer Science and Engineering from Kuvempu University, Shimoga in the year 1994 and M.Tech (Honors) Computer Science Engineering From Dr. M.G.R. Educational and Research Insititute , India. She is currently working as Lecturer in the Department of Computer Technology, MIT Campus, Anna University, Chennai, India. Her present research interests include, MANET, sensor network, mobile computing and network security.

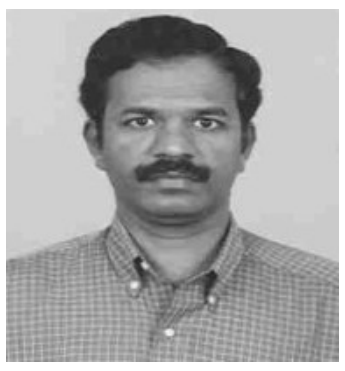

**Dr.D.Sridharan** received his B.Tech. degree and M.E.degree in Electronics Engineering from Madras Institute of Technology, Anna University in the years 1991 and 1993 respectively. He got his Ph.D degree in the Faculty of Information and Communication Engineering, Anna University in 2005. He is currently working as Assistant Professor in the Department of Electronics and Communication Engineering, CEG Campus, Anna University, Chennai, India. He was awarded the Young Scientist Research Fellowship by SERC of Department of Science and Technology, Government of India. His present research interests include Internet Technology, Network Security, Distributed Computing and Wireless Sensor Networks.